\documentstyle[amssymb,11pt,epsfig]{article}
\input{tcilatex}

\begin{document}

\title{Polarized Parton Distributions in the Valon Model}
\author{S. Atashbar Tehrani\thanks{%
atashbar@cic.aut.ac.ir} $^{(a)}$ , Ali N. Khorramian\thanks{%
khorramiana@theory.ipm.ac.ir} $^{(b,c)}$ \and \vspace{0.15in}\vspace{0.07in} \\
%EndAName
$^{(a)}$Physics Department, Persian Gulf University 75168, Boushehr, Iran\\
$^{(b)}$ Institute for Studies in Theoretical Physics and Mathematics (IPM)\\
P.O.Box 19395-5531, Tehran, Iran \\
$^{(c)}$ Physics Department, Semnan University, Semnan, Iran \\
}
\date{\today}
\maketitle

\begin{abstract}
We calculated the polarized parton distributions in leading order
approximation in QCD from valon model. Polarized structure
function of the nucleon are analyzed in the polarized valon model
in which a nucleon is assumed to be a bound state of three
polarized valence quark cluster(valons). It is shown that the
results of the model calculation agree rather well with the SMC
and available experimental data . Our results for $\Gamma _{1}$ of
hadron and deuteron  are in good agreement with experimental data
for some value of $Q^{2}$.
\end{abstract}

%\maketitle

\section{INTRODUCTION}

The idea of quark cluster is not new. In valon model, the hadron
is envisaged as a bound state of valence quark cluster called
valon \cite{1,2}. For example, the bound state of proton consists
of two ``up'' and one ``down'' valon. These valons thus bear the
quantum numbers of respective valence quarks. Hwa \cite{1} found
evidence in the deep inelastic neutrino scattering data that
suggested their existence and applied it to a variety of
phenomena. Hwa \cite{2} had also successfully formulated a
treatment of the low-$p_{T}$reactions based on the structural
analysis of the valons. Arash \cite{3} has recently applied the
valon model to study the contributions of the constituents of the
proton to its spin in deep inelastic scattering (DIS). Results of
the model calculation agree well with the European Muon
Collaboration (EMC) results \cite{4}. Arash adopt the valon point
of view and calculate various contributions to the spin of the
proton.

The valon description of the nucleon was developed by Hwa
\cite{5,6} some years ago and was successfully used \cite{3} to
describe the contribution of constituents of the proton to its
spin in order to understand the so-called {\it spin crisis.} In
this paper, we investigate the calculation of polarized parton
density of the nucleon using an improved model and apply the model
to the study of the spin of nucleon. In Sec. 2, polarized valon
distributions are derived. In this section we calculate polarized
valon distributions as a function of $y$ from unpolarized valon
distributions. The moment of polarized parton distributions and
using inverse Mellin transformation are discussed in Sec.3. In
Sec. 4, we examine the polarized structure function, $xg_{1}$, for
proton, neutron and deuteron and compare with the experimental
data. Also our prediction for the first moment of these structure
functions presents in this section too.

\section{UNPOLARIZED AND POLARIZED VALON DISTRIBUTIONS}

The polarized valon is defined to be a dressed polarized valence
quark in QCD with the cloud of polarized gluons and sea quarks
which can be resolved by high $Q^{2}$ probes. In a scattering
process the virtual emission and absorption of gluons in a valon
become bremsstrahlung and pair creation. The polarized structure
function of a valon is determined by gluon bremsstrahlung and pair
creation in QCD. At sufficiently low value of $Q^{2}$ the internal
polarized structure of a valon cannot be resolved and hence it
behaves as a polarized valence quark.

To facilitate the phenomenological analysis Hwa \cite{5} assumed a
simple form for the exclusive valon distribution in unpolarized
proton which is
\begin{equation}
G_{UUD/p}(y_{1},y_{2},y_{3})=g_{p}(y_{1}y_{2})^{\alpha
}y_{3}^{\beta }\delta (y_{1}+y_{2}+y_{3}-1)
\end{equation}
$y_{i}$ is the momentum fraction of the $i$th valon , and $g_p$ is
\[
g_{p}=[B(\alpha +1,\beta +1)B(\alpha +1,\alpha +\beta +2)]^{-1}
\]
where $B(m,n)$ is the beta function. The  valon distributions are
obtained by integration
\begin{eqnarray}
G_{U/p}(y) &=&\int dy_{2}\int dy_{3}G_{UUD/p}(y,y_{2},y_{3}) \\
&=&g_{p}B(\alpha +1,\beta +1)y^{\alpha }(1-y)^{\alpha +\beta +1}
\nonumber
\end{eqnarray}
\begin{eqnarray}
G_{D/p}(y) &=&\int dy_{1}\int dy_{2}G_{UUD/p}(y_{1},y_{2},y) \\
&=&g_{p}B(\alpha +1,\alpha +1)y^{\beta }(1-y)^{2\alpha +1}
\nonumber
\end{eqnarray}
Recently, R.C. Hwa \cite{7} have recalculated the unpolarized
valon distribution in proton with a new set of parameters. The new
values of $\alpha $ , $\beta $ are found to be $\alpha=1.76 $ and
$\beta =1.05 $.
 An alternative analysis of polarized
structure function has been performed by Gehrmann and
Stirling\cite {8}. This was done in the same spirit as the
unpolarized analysis \cite{9}, namely the polarized valon
distribution chose to be of the general form

\[
\delta G_{U/p}(y)=\delta F _{U}(y)\times G_{U/p}(y)
\]

\begin{equation}
\delta G_{D/p}(y)=\delta F _{D}(y)\times G_{D/p}(y)
\end{equation}

where $\delta \digamma _{j}(y)$ is define:

\begin{equation}
\delta F _{j}(y)=N_{j}y^{\alpha _{j}}(1-y)^{\beta _{j}}(1+\gamma
_{j}y+\delta _{j}y^{0.5})
\end{equation}

where the subscript $j$ refer to $U,D$, and $G_{\frac{v}{p}}(y)$
is defined in Eqs. (2,3). In order to write down the polarized
valon distribution in a proton we assume that at some $Q=Q_{0}$ a
valon behaves as a polarized valence quark. The polarized quark
distribution in a proton at a starting scale of $Q_{0}^{2}= 0.26
GeV^{2}$ is parameterized by GRV group \cite{10}. We take that
parameterization and extract all of the parameters as defined in
Eq.\-(5). The corresponding parameters are given in Table I. In
Figure (1) we present unpolarized and polarized valon
distributions as a function of $y$.

\section{POLARIZED PARTON DISTRIBUTIONS}

We now go to the $n$-moment space and define the moment of polarized valon
distribution $\Delta M_{v/p}(n)$ as:

\begin{equation}
\Delta M_{v/p}(n)\equiv \int_{0}^{1}y^{n-1}\delta G_{v/p}(y)dy
\end{equation}
the subscription $v$ stands for kind of $U,D$ polarized valon. For
the moments of polarized singlet and non singlet distributions we
shall use, the leading order solutions of the renormalization
group equation in QCD. They can be expressed entirely in terms of
the evolution parameter $s$
\begin{equation}
s=\ln \frac{\ln Q^{2}/\Lambda ^{2}}{\ln Q_{0}^{2}/\Lambda ^{2}}
\end{equation}

where $Q_{0},\Lambda$ are scale parameters to be determined by
experiments. From the theoretical standpoint both $Q_{0}$ and
$\Lambda$, depend on the order of the moments $n.$ In this work we
have assumed that they are independent of n and we fixed it by
$Q_{0}^{2}=0.26 GeV^{2}$ and $ \Lambda=0.232 GeV$.

The moments of singlet and non singlet polarized are
\begin{equation}
\Delta M^{NS}(n,Q^{2})=\exp (-\delta d_{NS}^{(0)n}s)
\end{equation}

\begin{equation}
\Delta M^{S}(n,Q^{2})=\frac{1}{2}(1+\delta \rho )\exp (-\delta d_{+}s)+\frac{%
1}{2}(1-\delta \rho )\exp (-\delta d_{-}s)
\end{equation}
where $\delta \rho $ and other associated parameters are as follows:

\begin{eqnarray}
\delta \rho &=&\frac{\delta d_{NS}^{(0)n}-\delta d_{gg}^{(0)n}+4f\delta
d_{qg}^{(0)n}}{\Delta }  \nonumber \\
\Delta &=&\delta d_{+}-\delta d_{-}=\sqrt{(\delta d_{NS}^{(0)n}-\delta
d_{gg}^{(0)n})^{2}+8f\delta d_{qg}^{(0)n}\delta d_{gq}^{(0)n}}  \nonumber \\
\delta d_{\pm } &=&\frac{1}{2}(\delta d_{NS}^{(0)n}+\delta d_{gg}^{(0)n}\pm
\Delta ) \\
b &=&\frac{33-2f}{12\pi }  \nonumber
\end{eqnarray}

and also the anomalous dimensions $\delta d_{ij}^{(0)n}$ are simply the n-th
moment of polarized LO splitting function are given by \cite{11} and $f$ is
the number of active flavors.

\[
\delta d_{NS}^{(0)n}=\frac{1}{3\pi b}[-3-\frac{2}{n(n+1)}+4S_{1}(n)]
\]
\[
\delta d_{qg}^{(0)n}=\frac{-1}{4\pi b}\frac{n-1}{n(n+1)}
\]
\[
\delta d_{gq}^{(0)n}=\frac{-2}{\pi b}\frac{1}{3}\frac{n+2}{n(n+1)}
\]

\begin{equation}
\delta d_{gg}^{(0)n}=-\frac{1}{2\pi b}\{3[\frac{11}{6}+\frac{4}{n(n+1)}%
-2S_{1}(n)]-\frac{2}{3}\frac{f}{2}\}.
\end{equation}

Here $S_{1}(n)=\sum^{n}_{j=1} \frac{1}{j}=\psi (n+1)+\gamma _{E},\psi (n)=%
\frac{\Gamma ^{\prime }(n)}{\Gamma (n)}$ and $\gamma _{E}=0.577216.$

The moment of polarized $u$ and $d$-valence quark in a proton are:

\[
\Delta M_{u_{v}}(n,s)=2\Delta M^{NS}(n,s)\Delta M_{U/p}
\]

\begin{equation}
\Delta M_{d_{v}}(n,s)=\Delta M^{NS}(n,s)\Delta M_{D/p}
\end{equation}

and the moment of polarized $\Sigma $ and gluon distributions are as follows:

\[
\Delta M_{\Sigma }(n,s)=\Delta M^{s}(n,s)(2 \Delta M_{U/p}+\Delta
M_{D/p})
\]

\begin{equation}
\Delta M_{g}(n,s)=\Delta M_{gq}(n,s)(2 \Delta M_{U/p}+\Delta
M_{D/p})
\end{equation}

where $\Delta M_{gq}(n,s)$ is the quark-to-gluon evolution
function give by:
\begin{eqnarray}
\Delta M_{gq}(n,s) &=&\frac{(\delta d_{+}-\delta d_{NS}^{(0)n})}{\Delta }(1-%
\frac{\delta d_{-}-\delta d_{NS}^{(0)n}}{2f\delta d_{qg}^{(0)n}})\exp -d_{+}s
\nonumber \\
&&-\frac{(\delta d_{-}-\delta d_{NS}^{(0)n})}{\Delta }(1-\frac{\delta
d_{+}-\delta d_{NS}^{(0)n}}{2f\delta d_{qg}^{(0)n}})\exp -d_{-}s
\end{eqnarray}

In Figure (2) we present these moments as a function of $n$ for some values
of $Q^{2}$. To obtain the $x-$dependence of structure functions and parton
distributions, usually required for practical purposes, from the above $n-$%
dependent exact analytical solutions in Mellin-moment space, one has to
perform a numerical integral in order to invert the Mellin-transformation in
(8) according to

\begin{equation}
\delta f(x,Q^{2})=\frac{1}{\pi }\int_{0}^{\infty }dz Im[e^{i\phi
}x^{-c-ze^{i\phi }}\Delta f^{n=c+ze^{i\phi }}(Q^{2})]
\end{equation}

where the contour of integration, and thus the value of $c$, has to lie to
the right of all singularities of $\Delta f^{n}(Q^{2})$ in the complex $n$%
-plane, i.e., $c\geq 0$ since according to Eq. (13) the dominant
pole of all $\delta d_{ij}^{n}$ is located at $n=0$. For all
practical purpose one may choose $c\simeq 1,\phi =135^{\circ }$
and an upper limit of integration, for any $Q^{2}$ , of about
$5+10/\ln x^{-1}$, instead of $\infty $, which suffices guarantee
and stable numerical results \cite{12,13}. In figure (3) our
predictions for $x-$dependence spin densities for quarks and
gluons at $Q^{2}=2, 3, 5, 10$ $GeV^{2}$ in a proton are shown.

Furthermore, in the polarized parton distribution we assume an unbroken $%
SU(3)_{f}$ symmetric sea,
\begin{equation}
\delta \overline{q}(x,Q^{2})\equiv \delta \overline{u}=\delta u_{sea}=\delta
\overline{d}=\delta d_{sea}=\delta \overline{s}=\delta s_{sea}
\end{equation}

Note that the proton include partons with positive and negative helicity
which carry of fraction $x$ of the proton momentum and to whom one can
associate quark densities $q_{+}(x,Q^{2})$ and $q_{-}(x,Q^{2})$. The
difference

\begin{equation}
\delta q(x,Q^{2})=q_{+}(x,Q^{2})-q_{-}(x,Q^{2})
\end{equation}

measures how much the parton of flavor $q$ ``remembers'' of its parent
parton polarization. Similarly, one may define

\begin{equation}
\delta \overline{q}(x,Q^{2})=\bar{q}_{+}(x,Q^{2})-\bar{q}_{-}(x,Q^{2})
\end{equation}

for antiquarks.

\section{POLARIZED STRUCTURE FUNCTION, $\bf g_1$}

We used polarized parton distribution from valon model for extracting
structure function $g_{1}$. Within this model according Sec. (3), we
calculated all of the polarized parton distributions in a proton. Now it is
very easy to used these results for calculating $g_{1}^{p}$. In the quark
model, to leading order (LO) in QCD, $g_{1}^{p}$ can be written as a linear
combination of $\delta q$ and $\delta \overline{q}$ \cite{11,14},

\begin{equation}
g_{1}^{p}(x,Q^{2})=\frac{1}{2}\sum_{q} e_{q}^{2}[\delta q(x,Q^{2})+\delta
\overline{q}(x,Q^{2})]
\end{equation}

where $e_{q}$ are the electric charges of the (light) quark-flavors $q=u,d,s$%
.

Furthermore, Eq. (19) can be decomposed into a flavor nonsinglet
(NS) and singlet (S) component

\begin{equation}
g_{1}^{p}(x,Q^{2})=g_{1,NS}(x,Q^{2})+g_{1,S}(x,Q^{2})
\end{equation}

where

\begin{equation}
g_{1,NS}(x,Q^{2})=\frac{1}{2}\sum_{q} (e_{q}^{2}-\langle e^{2}\rangle
)(\delta q+\delta \overline{q})
\end{equation}

with $\langle e^{2}\rangle =\frac{1}{f}\sum_{q} e_{q}^{2}$ (e.g. for f=3
light $u,d,s$ flavors $\langle e^{2}\rangle =\frac{2}{9}$)and

\begin{equation}
g_{1,S}(x,Q^{2})=\frac{1}{2}\langle e^{2}\rangle \sum_{q}(\delta q+\delta
\overline{q})\equiv \frac{1}{2}\langle e^{2}\rangle \delta \Sigma .
\end{equation}

where the sum usually runs over the light quark-flavors $q=u,d,s$, since the
heavy quark contributions (c, b,...) have preferably to be calculated
perturbatively from the intrinsic light quark ($u,d,s$) and gluon ($g$)
partonic-constituent of the nucleon.

We are now in a position to present the results for the proton
polarized structure function, $g_{1}^{p}$. In Eq. (19) we
presented polarized proton structure function in a proton. In
figure (4) the results are shown at some values of $Q^{2}$. The
data points are from \cite{15}. For the neutron structure function
one only needs to interchange $u_{v}$ and $d_{v}$ in the Eq. (19).
For deuteron it is given by:
\begin{equation}
g_{1}^{d}(x,
Q^{2})=\frac{1}{2}[g_{1}^{p}(x,Q^{2})+g_{1}^{n}(x,Q^{2})](1-\frac{3}{2}\omega_{D})
\end{equation}
where $\omega_{D} \approx 0.058$ accounts for the D-state
admixture in the deuteron wave function. Quantities of particular
significance are the first moment $(n=1)$ of
polarized parton distributions $\delta f(x,Q^{2}),$%
\begin{equation}
\Delta f(Q^{2})\equiv \int_{0}^{1}dx\delta f(x,Q^{2})\text{ \quad , \quad }%
f=q,\overline{q},g
\end{equation}

We calculated first moments (total polarizations) $\Delta f(Q^{2})$ of the
polarized parton distributions. These results shown in Table II.

According to Eqs. (17) and (18), $\Delta (q+\overline{q})$ is the
net number
of right-handed quarks of flavor $q$ inside a right-handed proton and thus $%
\frac{1}{2}\Delta \Sigma $, is a measure of how much all quark flavors
contribute to the spin of the proton. Similarly, $\Delta g$ represents the
total gluonic contribution to the spin of the nucleon.

Here we have defined the first moment of $g_{1}$ (Ellis-Jaffe sum
rule) by

\[
\Gamma _{1}^{p,n,d}(Q^{2})\equiv
\int_{0}^{1}dxg_{1}^{p,n,d}(x,Q^{2})
\]

Our values for $\Gamma _{1}^{p,n,d}$ represents in Table III,
which are in excellent agreement with the E143, E142, E154, SMC
results from Ref. [16-19].

Figure (5) shows our results for $xg_{1}^{n}$ at $Q^{2}=2,3,5$ $GeV^{2}$
which are in agreement with the results of experimental data from \cite
{15,17,18}.

In figure (6) we have shown $xg_{1}^{d}$ for some values of $Q^{2}$. The
experimental data are from \cite{15}.

Finally by this model it is possible to attribute numbers $\triangle \Sigma $
and $\Delta g$ to the quark and gluon spin content of the nucleons which
describe their total (integrated) contribution to the nucleon spin in the
following sense
\begin{equation}
\frac{1}{2}=\frac{1}{2}\Delta \Sigma(Q^2) +\Delta
g(Q^2)+L_{z}(Q^2)
\end{equation}

where on the left hand side we have the spin $(+\frac{1}{2})$ of a
polarized nucleon state and on the right hand side a decomposition
in terms of $\Delta \Sigma (Q^{2})=\Delta u+\Delta d+\Delta
s+\Delta \overline{u}+...$, $\Delta g(Q^{2})$ and the relative
orbital angular momentum $L_{z}(Q^{2})$ among all the quarks and
gluons. Unfortunately, the decomposition in Eq. (25) cannot
directly be measured in experiments. Instead various other
combinations of $\Delta \Sigma$ and $\Delta g$ appear in
experimental observables. Our calculation for study of nucleon
spin presented in Fig. (7).

\section{CONCLUSION}

We have presented a model in which the polarized parton density is
calculated. In this paper polarized valon distribution is derived
from unpolarized valon distribution. After define the moment of
polarized valon distribution and moments of all partons in proton,
polarized parton density in a hadron are available. The results
are used to evaluate the spin components of nucleon. It turns out
that are results about polarized structure function agree well
with all available experimental data on $g_{1}$ of proton ,neutron
and deuteron. Also our prediction for $\Gamma_{1}$ is good
agreement with experimental data.

\newpage
\textbf{Table Caption}\\

\begin{tabular}{cccccc}
\hline\hline $v$ & $N$ & $\alpha $ & $\beta $ & $\gamma $ & $\eta
$
\\ \hline
$U$ & 0.0475 & -1.003 & -0.998 & 0.653 & -0.067 \\
$D$ & -0.015 & -1.247 & -0.951 & 8.351 & -2.063 \\ \hline\hline
\end{tabular}
\\
\\
 \textbf{Table I.} The parameters defined in Eqs. (4,5) for $U$ and
$D$ polarized valon distribution.
\\
\\

\begin{tabular}{cccccc}
\hline\hline
$Q^{2}(GeV^{2})$ & $\Delta u_{v}$ & $\Delta d_{v}$ & $\Delta \Sigma $ & $%
\Delta \overline{q}$ & $\Delta g$ \\ \hline
2  & 0.609 & -0.306 & 0.329 & 0.004 & 0.876 \\
3  & 0.609 & -0.306 & 0.334 & 0.005 & 0.992 \\
5 & 0.609 & -0.306 & 0.341 & 0.006 & 1.139 \\
10 & 0.609 & -0.306 & 0.350 & 0.008 & 1.334 \\ \hline\hline
\end{tabular}
\\
\\
\textbf{Table II.} The first moments of polarized parton
distributions, $\Delta u_{v}$, $\Delta d_{v}$, $\Delta \Sigma $,
$\Delta \overline{q}$ and $\Delta g$ for some value of $Q^{2}$.
\\
\\
\\
\\
\begin{tabular}{cccc}
\hline\hline $Q^{2}(GeV^{2})$ & $\Gamma _{1}^{p}$ & $\Gamma
_{1}^{n}$ & $\Gamma _{1}^{d}$
\\ \hline
2 & 0.1212 & -0.0313 & 0.0412 \\
3 & 0.1219 & -0.0307 & 0.0417 \\
5 & 0.1225 & -0.0299 & 0.0424 \\
10 & 0.1236 & -0.0289 & 0.0433 \\ \hline\hline
\end{tabular}
\\
\\
\textbf{Table III.} The first moments of proton, neutron, and
deuteron polarized structure function for $Q^{2}=2, 3, 5, 10$
$GeV^{2}$.

\newpage

\begin{figure}[tbh]
\centerline{\includegraphics[width=0.51\textwidth]{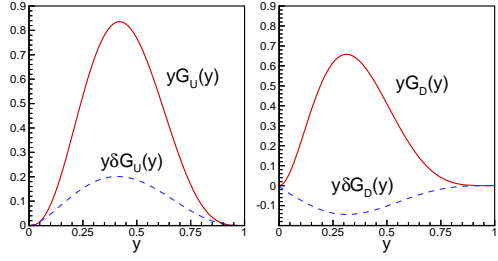}}
\caption{The unpolarized and polarized valon distributions as a
function of $y$.}
\end{figure}

\begin{figure}[tbh]
\centerline{\includegraphics[width=0.6\textwidth]{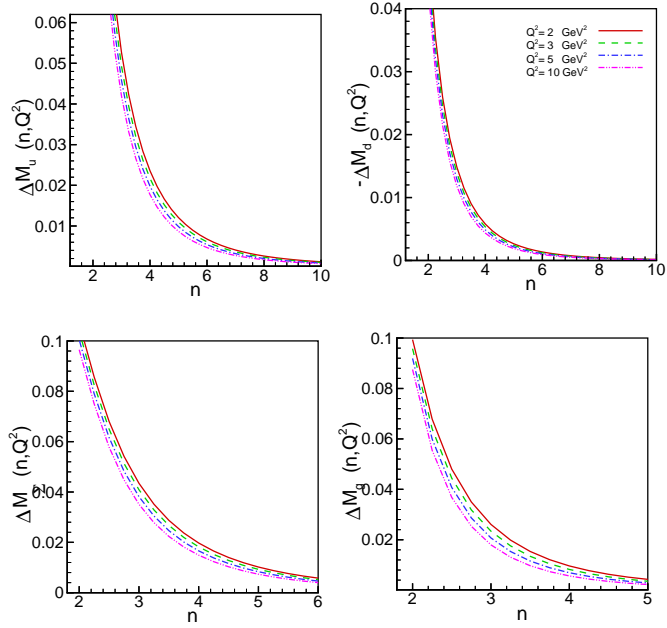}}
\caption{Moments of polarized parton distributions as a function
of $n$ for some values of $Q^{2}$.}
\end{figure}

\begin{figure}[tbh]
\centerline{\includegraphics[width=0.5\textwidth]{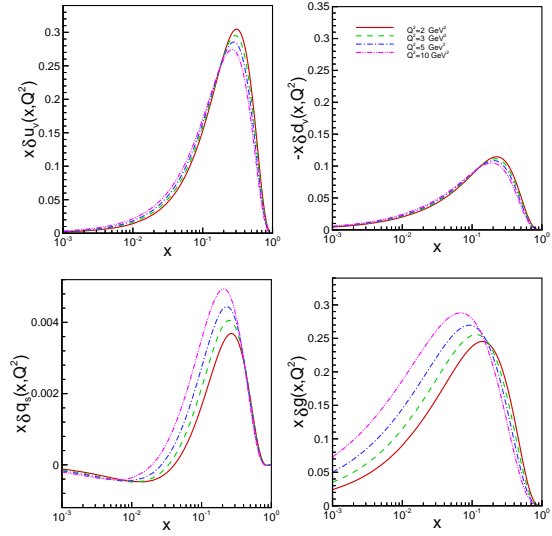}}
\caption{Polarized parton distributions in proton at $Q^{2}=2, 3,
5, 10GeV^{2}$ as a function of $x$.}
\end{figure}

\begin{figure}[tbh]
\centerline{\includegraphics[width=0.5\textwidth]{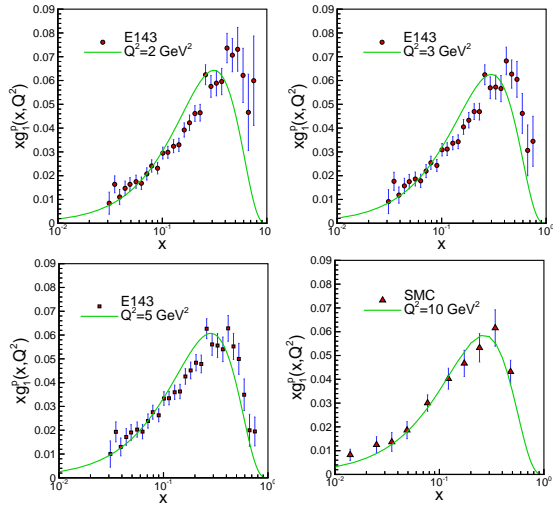}}
\caption{The compares of our predictions for $xg_{1}^{p}(x,Q^{2})$
versus $x$ from valon model with the measured structure functions
in experiments on proton target. Data points are from \cite{15}. }
\end{figure}

\begin{figure}[tbh]
\centerline{\includegraphics[width=0.5\textwidth]{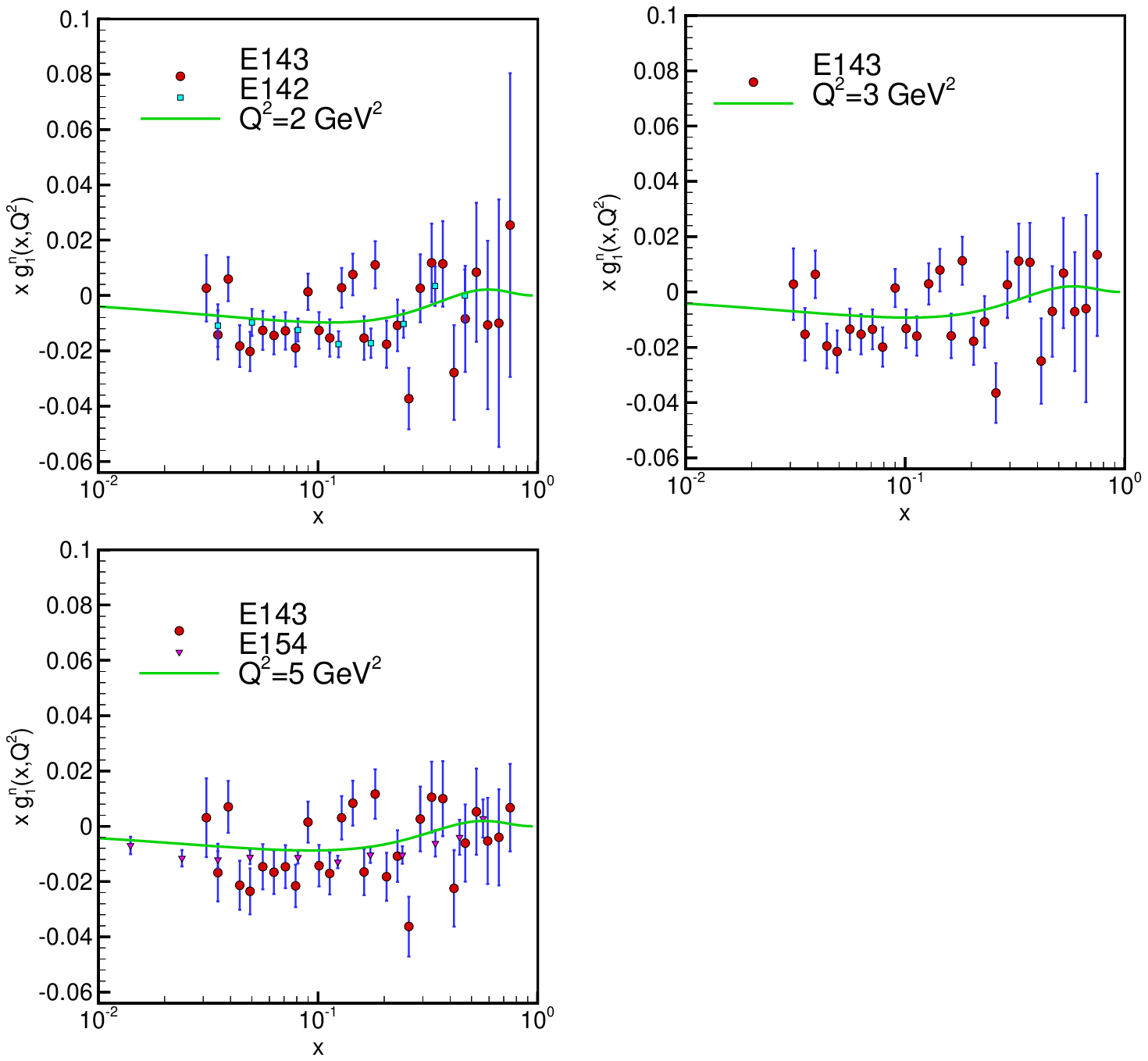}}
\caption{Plots of neutron polarized structure function against $x$ at $%
Q^{2}=2,3,5$ $GeV^{2}$. Data points are from\cite{15,17,18}. }
\end{figure}

\begin{figure}[tbh]
\centerline{\includegraphics[width=0.5\textwidth]{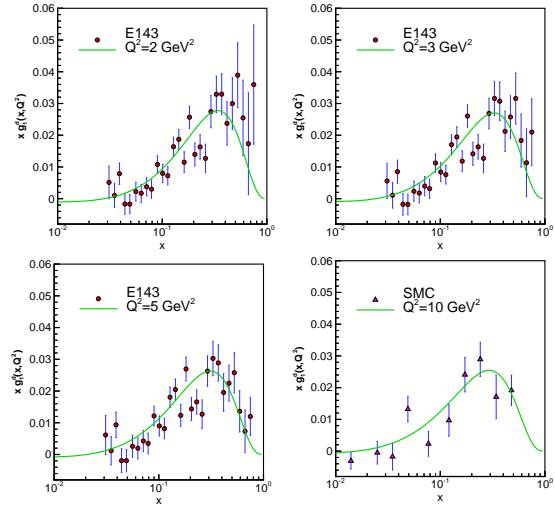}}
\caption{Plots of deuteron polarized structure function against $x$ at $%
Q^{2}=2,3,5$ and $10$ $GeV^{2}$. Data points are from \cite{15}. }
\end{figure}

\begin{figure}[tbh]
\centerline{\includegraphics[width=0.5\textwidth]{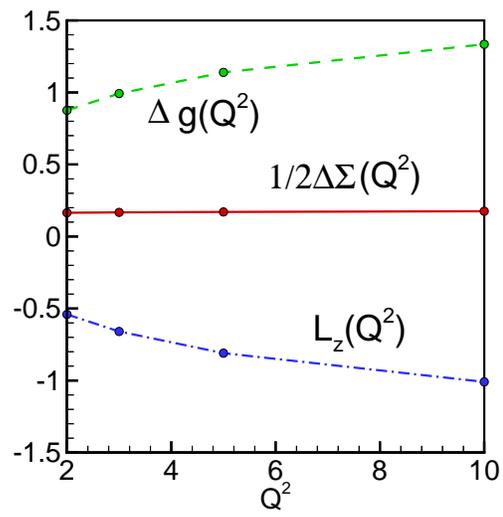}}
\caption{The variation of $\frac{1}{2}\Delta \Sigma$, $\Delta g$
and $L_{z}$ as a function of $Q^{2}$. }
\end{figure}

\end{document}